\documentclass[showpacs,aps,graphicx,twocolumn]{revtex4}
\usepackage{graphicx}

\begin{document}
\title{Quantum secure direct communication network
 with superdense coding and decoy photons}
\author{ Fu-Guo Deng,$^{1,2,3}$\footnote{ E-mail: fgdeng@bnu.edu.cn} Xi-Han Li,$^{1,2}$ Chun-Yan Li,$^{1,2}$  Ping Zhou,$^{1,2}$
and Hong-Yu Zhou$^{1,2,3}$}
\address{$^1$ Key Laboratory of Beam Technology and Material
Modification of Ministry of Education, Beijing Normal University,
Beijing 100875,
People's Republic of China\\
$^2$ Institute of Low Energy Nuclear Physics, and Department of
Material Science and Engineering, Beijing Normal University,
Beijing 100875, People's Republic of China\\
$^3$ Beijing Radiation Center, Beijing 100875,  People's Republic
of China}
\date{\today }

\begin{abstract}
A quantum secure direct communication network scheme is proposed
with quantum superdense coding and decoy photons. The servers on a
passive optical network prepare and measure the quantum signal,
i.e., a sequence of the $d$-dimensional Bell states. After
confirming the security of the photons received from the receiver,
the sender codes his secret message on them directly. For preventing
a dishonest server from eavesdropping, some decoy photons prepared
by measuring one photon in the Bell states are used to replace some
original photons. One of the users on the network can communicate
any other one. This scheme has the advantage of high capacity, and
it is more convenient than others as only a sequence of photons is
transmitted in quantum line.

\end{abstract}
 \maketitle

\section{introduction}

Recently, a novel branch of quantum communication, quantum secure
direct communication (QSDC) was proposed and actively pursued by
some groups
\cite{beige,two-step,QOTP,Wangc,Wangc2,bf,caiA,zhangzj,yan,gaot,zhangs,lixhpra,wangj,leepra,song}.
Different from QKD, the goal of QSDC is that the two parties of
communication, say Bob and Charlie can exchange the secret message
directly without generating a private key first and then encrypting
the message. The scheme proposed by Beige et al. \cite{beige} can be
used to transmitted the secret message after the transmission of an
additional classical bit for each qubit. Yan and Zhang \cite{yan}
proposed a QSDC protocol with quantum teleportation. Gao, Yan and
Wang \cite{gaot} introduced a QSDC protocol with quantum
entanglement swapping. Man, Zhang and Li \cite{zhangzj} presented a
QSDC protocol with entanglement swapping too. Zhu et al.
\cite{zhangs} proposed a QSDC protocol with Einstein-Podolsky-Rosen
(EPR) pairs based on the encryption on the order of the transmission
of the particles, similar to that in the QKD protocol \cite{CORE}.
Wang, Zhang and Tang \cite{wangj} revised this QSDC protocol with
single photons, and Li et al. \cite{lixhpra} improved its security
against Trojan horse attack. The secret message  in all these QSDC
protocols can be read out by the receiver only after at least one
bit of classical information is exchanged for each qubit. More
accurately, they are some deterministic secure quantum communication
protocols pointed out by Li et al. \cite{LIDSQC}, and are very
useful for deterministic communication when the noise in a quantum
line is not very low. Bostr\"{o}m and Felbinger \cite{bf} proposed
an interesting ping-pong QSDC following some ideas in quantum dense
coding \cite{BW} with an EPR pair even though W\'ojcik
\cite{attack1}  proved that the ping-pong protocol is insecure for
direct communication if there are losses in a practical quantum
channel and we \cite{dengattack} showed that it can be eavesdropping
with multi-photon fake signal fully if there is noise in the quantum
channel. Cai and Li \cite{caiA} improved the capacity of ping-pong
protocol. We put forward a two-step QSDC protocol \cite{two-step}
with EPR pairs transmitted in block and another one based on a
sequence of polarized single photons \cite{QOTP}. Wang et al.
introduced a QSDC protocol with high-dimensional quantum superdense
coding \cite{Wangc} and another one with Greenberger-Horne-Zeilinger
(GHZ) states \cite{Wangc2}. The good nature of the QSDC schemes
\cite{two-step,QOTP,Wangc} with quantum data block is that the
participants can perform quantum privacy amplification
\cite{deutschqpa,QPA} on the unknown states to improve their
security in a noisy channel. Lee, Lim and Yan \cite{leepra}
introduced a protocol for controlled secure direct communication
with GHZ states. Cao and Song proposed a QSDC protocol with W state
and teleportation \cite{song}.

A practical quantum communication requires that any one on a passive
optical network can communicate another authorized user, similar to
a classical communication network, such as the world wide web (i.e.,
www) and the classical telephone network. Usually, there are some
servers (the number of the servers is much less than that of the
users), say Alice$_s$ who provide the service for preparing and
measuring the quantum signal for the legitimate users on a passive
optical network, which will reduce the requirements on the users'
devices for secure communication (not the servers) largely, same as
the classical communication. By far, there are few any-to-any QSDC
network schemes \cite{lxhnetwork} although there are some
point-to-point QSDC schemes
\cite{beige,two-step,QOTP,Wangc,Wangc2,bf,caiA,zhangzj,yan,gaot,zhangs,lixhpra,wangj,leepra}
existing, which is different from QKD \cite{Phoenix,DLMXL,LZWD}.
Moreover, the existing QSDC point-to-point schemes
\cite{beige,QOTP,Wangc,Wangc2,bf,caiA,zhangzj,yan,gaot,zhangs,lixhpra,wangj,leepra}
cannot be used directly to accomplish the task in a QSDC network
except for the two-step QSDC protocol \cite{two-step}. The reason is
that a dishonest server can steal some information without being
detected. In order to be practical and secure, the security of a
QSDC network must be guaranteed against a dishonest server as the
number of server is larger than one. For example, when the
point-to-point QSDC schemes in Refs
\cite{QOTP,Wangc,Wangc2,bf,caiA,zhangzj,yan,gaot} are used for QSDC
network directly, the server Alice can steal the information by
using the intercepting-resending attack without disturbing the
quantum signal as she prepares it and knows its original state
\cite{note}. The QSDC network protocol in Ref. \cite{lxhnetwork} is
composed of three two-step point-to-point QSDC schemes. All the
particles in the EPR pairs should be transmitted from the server to
the sender and then to the receiver before they run back to the
server.  In 2005, Gao, Yan and Wang \cite{yannetwork} proposed a
QSDC protocol for a central party and his agents. The quantum
information can only be transmitted from the agents to the center,
not any one to any one on the network.

In this paper, we will introduce a QSDC network scheme following
some ideas in quantum superdense coding \cite{BW,superdense} with an
ordered $N$ EPR photon pairs. One authorized user can communicate
any one on the network securely with the capability of measuring
$d$-dimensional single-photon states. In this scheme, only one
particle in each EPR pair is transmitted, which will make it more
convenient than that in Ref. \cite{lxhnetwork}. The decoy photons
will ensure its security.

\section{the QSDC network with superdense coding}

\begin{figure}[!h]
\begin{center}
\includegraphics[width=8cm,angle=0]{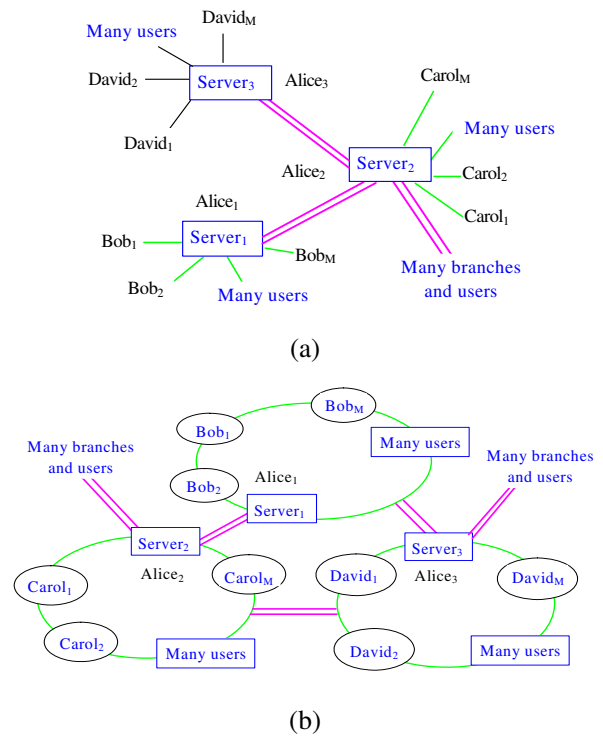}
\caption{ The topological structure of the network, similar to those
in Refs. \cite{Phoenix,DLMXL,LZWD}. (a) star-configuration network;
(b) loop-configuration network. }
\end{center}\label{cell}
\end{figure}

In general, the structure of a QSDC network is as the same as that
for quantum key distribution \cite{Phoenix,DLMXL,LZWD}. There are
two kinds of topological structures for QSDC networks, a loop one or
a star one, shown in Fig.1. The QSDC network can work similar to QKD
network \cite{DLMXL}. That is, the classical message transmitted by
any of the authorized users can be gathered by the server Alice
through the identity authentication in a different time slot, and
the sender Bob and the receiver Charlie can distinguish their
identities each other. This task  can be accomplished with a
classical channel in which the classical message can be
eavesdropping, not be altered \cite{Gisinqkd}. The server Alice
provides the service for preparing and measuring the quantum signal,
which will reduce the requirements on the users' devices largely.
For each request of the users, the server analyzes it and then
performs a relevant operation, for example, connecting the user to
another one on the network or sending a sequence of photons to the
user. If two users do not exist in a branch of the network, they can
agree that the server of the branch with the receiver provides the
service for preparing and measuring the quantum signal, and the
other servers only provide the service for connecting the quantum
line for these two users in some a time slot. In this way, the
principle of a QSDC network is explicit if we can describe clearly
that of its subsystem, shown in Fig.2.  We only discuss the
principle for QSDC with the subsystem of the network below.

\begin{figure}[!h]
\begin{center}
\includegraphics[width=5.5cm,angle=0]{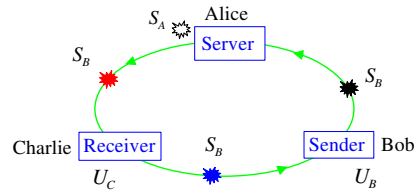}
\caption{ The subsystem of networks in this QSDC network. The server
Alice always keeps the sequence $S_A$ and sends $S_B$ firs to the
receiver of the secret message, Charlie who encrypts them with the
operations $U_C$. }
\end{center}\label{cell}
\end{figure}

A symmetric $d$-dimensional EPR photon pair is in one of the
Bell-basis states as following \cite{BW,superdense}
\begin{equation}
\vert \Psi_{nm}\rangle_{AB} =\sum_{j=0}^{d-1}
\frac{1}{\sqrt{d}}e^{2\pi ijn/d}\vert j\rangle_A \otimes \vert
j+m\;{\rm mod} \; d \rangle_B
\end{equation}
where $n,m=0,1,...,d-1$, and $A$ and $B$ are the two photons in an
EPR pair. The unitary operations
\begin{equation}
U_{nm} =\sum_{j=0}^{d-1} e^{2\pi ijn/d}\vert j+m\;{\rm mod} \; d
\rangle \langle j\vert
\end{equation}
can transform the Bell-basis state
\begin{equation}
\vert \Psi_{00}\rangle_{AB} =\sum_{j} \frac{1}{\sqrt{d}}\vert
j\rangle_A \otimes \vert j \rangle_B
\end{equation}
into the Bell-basis state $\vert \Psi_{nm}\rangle_{AB}$, i.e., $(I^A
\otimes U_{nm}^B)\vert \Psi_{00}\rangle_{AB}=\vert
\Psi_{nm}\rangle_{AB}$. Here $I^A$ is the identity matrix which
means doing nothing on the photon $A$, and $U_{nm}^B$ are only
operated on the photon $B$. The two sets of unbiased measuring bases
(MBs) can be chosen similar to those in a two-dimensional quantum
system, labeled as $Z_{d}$ and $X_{d}$. Let us define the $Z_{d}$
-MB which has $d$ eigenvectors as following:
\begin{eqnarray}
\left\vert  0 \right\rangle, \;\;\;\left\vert  1 \right\rangle,
\;\;\;\;\left\vert  2 \right\rangle \;\; \cdots, \;\;\;\;\left\vert
{d - 1} \right\rangle.
\end{eqnarray}
The $d$ eigenvectors of the $X_{d}$ -MB can be described as
\begin{eqnarray}
\vert 0\rangle_x&=&\frac{1}{{\sqrt d }}\left( {\left\vert  0
\right\rangle + \vert 1\rangle \;\; + \cdots \;\; + \left\vert
{d-1}\right\rangle }\right),\;\nonumber \\
\vert 1\rangle_x&=&\frac{1}{{\sqrt d }}\left({\left\vert  0
\right\rangle + e^{{\textstyle{{2\pi i} \over d}}} \left\vert  1
\right\rangle + \cdots
+ e^{{\textstyle{{(d-1)2\pi i} \over d}}} \left\vert  {d-1} \right\rangle} \right),\; \nonumber\\
\vert 2\rangle_x&=&\frac{1}{{\sqrt d }}\left({\left\vert  0
\right\rangle + e ^{{\textstyle{{4\pi i} \over d}}} \left\vert 1
\right\rangle + \cdots + e^{{\textstyle{{(d-1)4\pi i} \over d}}}
\left\vert  {d-1} \right\rangle }\right),\nonumber\\
&&\cdots \cdots \cdots \cdots \cdots \cdots \nonumber\\
\vert d-1\rangle_x&=&\frac{1}{{\sqrt d }}(\left\vert  0
\right\rangle + e ^{{\textstyle{{2(d-1)\pi i} \over d}}}
\left\vert 1 \right\rangle  + e ^{{\textstyle{{2\times 2(d-1)\pi
i} \over d}}}
\left\vert 2 \right\rangle + \cdots \nonumber\\
&& + e^{{\textstyle{{(d-1)\times 2(d-1)\pi i} \over d}}}
\left\vert {d-1} \right\rangle ).
\end{eqnarray}
The two vectors $\vert k\rangle$ and $\vert l\rangle_x$ coming from
two MBs satisfy the relation $\vert \langle k|l\rangle_x \vert
^2=\frac{1}{d}$,
 and the Hadamard ($H_d$) operation
\begin{eqnarray}
H_d  =\frac{1}{\sqrt{d}} \left( {\begin{array}{*{20}c}
   1 & 1 &  \cdots  & 1  \\
   1 & {e^{2\pi i/d} } &  \cdots  & {e^{(d-1)2\pi i/d} }  \\
   1 & {e^{4\pi i/d} } &  \cdots  & {e^{(d-1)4\pi i/d}  }\\
    \vdots  &  \vdots  &  \cdots  & \vdots  \\
   1 & {e^{2(d-1)\pi i/d} } &  \cdots  & {e^{(d-1)2(d-1)\pi i/d} }  \\
\end{array}} \right)\label{HD}
\end{eqnarray}
can realize the transformation between the states coming from the
$Z_{d}$ -MB and the $X_{d}$ -MB, i.e., $H_d\vert j\rangle=\vert
j\rangle_x$. Certainly, there are at most $d+1$ sets of unbiased
measuring bases for people to prepare and measure a $d$-dimensional
quantum system \cite{multimbs}. Let us assume that the users use $M$
MBs for preparing and measuring their quantum signals.

For the quantum communication, the server Alice prepares a sequence
of EPR pairs, an ordered $N$ $d$-dimensional photon pairs, in the
same state $\vert \Psi_{00}\rangle_{AB}$. She takes one photon (A)
from each entangled photon pair to make up an ordered partner photon
sequence, say the sequence $S_A$: $[P_1(A)$, $P_2(A)$, $\ldots$,
$P_N(A)]$, same as Ref. \cite{two-step}. Here the subscript numbers
are the orders of the $N$ EPR pairs. The remaining partner photons
in the photon pairs make up the other sequence, say $S_B$:
$[P_1(B)$, $P_2(B)$, $\ldots$, $P_N(B)]$. Here $P_i(A)P_i(B)$ are
the two photons in the $i$-th EPR pair. Different from the QKD
network schemes \cite{Phoenix,DLMXL,LZWD}, the server Alice first
sends the sequence $S_B$ to the receiver Charlie, and always keeps
the sequence $S_A$ by herself, depicted in Fig.1. Charlie chooses
one of the two modes, a small probability $p$ with the checking mode
and a large probability $1-p$ with the coding mode, for the photons
received. If he chooses the checking mode, Charlie requires Alice
perform the measurements on the correlated photons in the sequence
$S_A$ with $M$ MBs first, and then tell him the results. Charlie
measures the samples with the same MBs as those of Alice's and
analyzes the error rate of the samples. If the coding mode is
chosen, Charlie operates each photon in the sequence $S_B$ with one
of the $d\times d$ local unitary operations $\{U_{nm}\}$
($n,m=0,1,2,\dots, d-1$) chosen randomly, say $U_C$. For the
eavesdropping check of the transmission between Charlie and Bob,
Charlie picks up some samples distributing randomly in $S_B$ and
replaces them with the decoy photons which are prepared by Bob with
one of the $M$ MBs chosen randomly. He sends the sequence $S_B$ to
Bob. After the confirmation of the security of the transmission, Bob
codes his message on the sequence $S_B$ with the corresponding local
unitary operation $U_{ij}$, say $U_B$.  He sends the sequence $S_B$
to Alice who performs the  joint Bell-basis measurements on the
photon pairs and announces the results. Before his coding, Bob
chooses $k$ photons in $S_B$ and operates them with choosing
randomly one of the unitary operations $\{U_{nm}\}$, and uses them
for the eavesdropping check in the transmission from Bob to Alice.

After the security of the transmission of one batch of $N$ order
entangled photons is confirmed, the user codes the message on the
photons and then sends them to the server Alice. The operation $U_C$
done by the receiver Charlie is used to hide the message (i.e., the
operation $U_B$) coded by the sender Bob, and the decoy technique is
used to forbid the dishonest server to steal the information about
the operation $U_C$ freely. In this way, the QSDC network is
performed until the communication session is ended.

Now, let us describe the steps for the subsystem of our QSDC
network scheme in detail as follows.

(S1) The server Alice prepares a sequence of EPR photon pairs in
the same state $\vert \Psi_{00}\rangle_{AB}$, and she divides them
into two partner photon sequences $S_A$ and $S_B$. That is, she
takes the photon $A$ in each photon pair $\vert
\Psi_{00}\rangle_{AB}$ to form the sequence $S_A$, and the other
photons make up the sequence $S_B$. Alice sends the sequence $S_B$
to the receiver of the secret message, Charlie, and always keeps
the sequence $S_A$ at home.

(S2) After receiving $S_B$, Charlie checks the security of the
transmission by the following method. (a) Charlie chooses randomly
some photons from the sequence $S_B$ as the samples for
eavesdropping check. (b) Charlie requires Alice to measure the
correlated photons in $S_A$ with the $M$ MBs chosen randomly, and
tell him the results and her MBs. (c) Charlie takes the suitable
measurements on the corresponding photons in his samples with the
same MBs as those of Alice's. (d) Charlie analyzes the error rate by
comparing his results with Alice's to determine whether Eve is
monitoring the quantum line. This is called the first eavesdropping
check. If the error rate is very lower than the threshold
$\epsilon_t$, they continue the quantum communication to the next
step, otherwise they will discard the results and repeat the quantum
communication from the beginning.

(S3) Charlie encrypts each photon in $S_B$ by choosing randomly one
of the local unitary operations $\{U_{nm}\}$, say $U_C$, and then
sends the sequence $S_B$ to the sender of the secret message, Bob.
For the second eavesdropping check, Charlie chooses randomly some of
the photons in $S_B$ as the samples, and replaces them with the
decoy photons, say $s_{e1}$, prepared in advance with choosing
randomly one of the $M$ MBs before they are sent to the quantum
line. He keeps the secret about the positions and states of these
samples.

(S4) Bob and Charlie take the second eavesdropping check with some
samples. They can accomplish this task with the following method.
(a) Charlie tells Bob the positions of the decoy photons $s_{e1}$
and their states.  (b) Bob takes the suitable measurements on the
photons in $s_{e1}$ with the correlated MBs. (c) Bob compares his
results with those of Charlie's and determines whether the
transmission of the sequence $S_B$ from Charlie to Bob is secure. If
the transmission is secure, the quantum communication is continued
to the next step, otherwise it will be aborted and repeated from the
beginning.

(S5) Bob codes his secret message on the photons remained in $S_B$
with the corresponding local unitary operation $U_{ij}$, say $U_B$,
according to his secret message. He picks up some of the photons in
$S_B$ as the samples $s_{e2}$ for checking the security of the whole
quantum communication. That is, he performs randomly one of the
operations $\{U_{nm}\}$ on each sample. Bob sends the sequence $S_B$
to the server Alice.

(S6) Alice performs the joint $d$-dimensional Bell-basis
measurements on the EPR photon pairs and publishes the results, say
$U_A$.

(S7) Charlie and Bob check the security of the whole quantum
communication with the samples $s_{e2}$. If the error rate of the
samples is reasonably low, Charlie can read out the secret message
$U_B$ as $U_A= U_C  U_B$. Also, they can use some special techniques
to correct the errors in the secret message, similar to the two-step
protocol \cite{two-step}.

\section{Discussion and summary}

It is obvious that the present protocol is secure if the process of
the quantum communication between Charlie and Bob is secure.  As the
principle of the security in a quantum communication protocol
depends on the fact that an eavesdropper's action can be detected by
analyzing the error rate of the samples chosen randomly with
statistical theory. In essence, the procedure for analyzing the
error rate of the samples in this paper is same as that in the BB84
QKD protocol \cite{BB84} and its modified version, the
favored-measuring-basis QKD protocol which has been proposed by Lo
et al. \cite{ABC} and proven to be secure unconditionally. That is,
the authorized users choose randomly $M$ MBs to prepare and measure
the samples. So the security of this protocol is similar to those in
the BB84 protocol and its modified version with quantum privacy
amplification \cite{QPAE}. Moreover, it appears that
high-dimensional quantum communication protocols provide better
security than that obtainable with two-dimensional quantum systems,
as has been discussed in detail in Ref. \cite{BP} because the two
authorized users can use more than two sets of unbiased measuring
bases to check eavesdropping. Suppose that the two authorized users
use $M$ sets of unbiased MBs to check eavesdropping, the error rate
introduced by Eve's eavesdropping is in principle
$\epsilon_e=\frac{Md+1-M-d}{Md}$. If $d=3$ and $M=4$,
$\epsilon_e=50\%$ which is twice of that in BB84 QKD. That is, Eve's
action will introduce more errors in the results obtained by the two
authorized users with measurements on the decoy photons and then be
detected easily. Bechmann-Pasquinucci and Peres gave out the four
unbiased MBs for three-dimensional quantum system in Ref. \cite{BP}.
In this time, the MB $Z_3$ can be chosen as
\begin{equation}
\vert 0\rangle,\;\;\;\;  \vert 1\rangle, \;\;\;\;  \vert 2\rangle,
\end{equation}
and the MB $X_3$ as
\begin{eqnarray}
\vert x_{0}\rangle &=& \frac{1}{\sqrt{3}}(\vert 0\rangle+\vert
1\rangle+\vert 2\rangle),\\
\vert x_{1}\rangle &=& \frac{1}{\sqrt{3}}(\vert 0\rangle+ e^{2\pi
i/3}\vert 1\rangle+ e^{-2\pi i/3}\vert 2\rangle),\\
\vert x_{2}\rangle &=& \frac{1}{\sqrt{3}}(\vert 0\rangle+ e^{-2\pi
i/3}\vert 1\rangle+ e^{2\pi i/3}\vert 2\rangle).
\end{eqnarray}
The two other bases can be taken as
\begin{equation}
\frac{1}{\sqrt{3}}(e^{2\pi i/3}\vert 0\rangle+ \vert 1\rangle+ \vert
2\rangle) \,\, \rm{and \,\, cyclic \,\, permutation},
\end{equation}
and
\begin{equation} \frac{1}{\sqrt{3}}(e^{-2\pi i/3}\vert 0\rangle+
\vert 1\rangle+ \vert 2\rangle) \,\, \rm{and \,\, cyclic \,\,
permutation}.
\end{equation}
In a low noise quantum channel, the entanglement purification
technique \cite{hdpurification} for high-dimensional quantum system
can be used to reduce the information leaked about the operation
$U_C$.

It is worth pointing out the advantage that the sequence $S_B$ is
sent first to the receiver Charlie, not the sender Bob. The server
Alice cannot steal the secret message yet if she is dishonest. That
is, Charlie first encrypts the quantum states which will be used to
carry the secret message, by choosing randomly one of the local
unitary operations $\{U_{nm}\}$, and then sends them to the sender
Bob. The operations done by Charlie $U_C$ is equivalent to the
quantum key in quantum one-time-pad crypto-system
\cite{Gisinqkd,QOTP} if Alice or an eavesdropper does not eavesdrop
the quantum line between Charlie and Bob. On the other hand, it is
necessary for Charlie and Bob to exploit the decoy technique to
forbid the server Alice to eavesdrop the quantum line between
Charlie and Bob freely and fully. If there are no decoy photons in
$S_B$, Alice can read out the operation $U_C$ fully by intercepting
the sequence $S_B$ when it is transmitted from Charlie to Bob and
take the Bell-basis measurements on the $N$ order EPR photon pairs.
Fortunately, the decoy photons inserted by Charlie can prevent Alice
from stealing the information about the quantum key $U_C$ as Alice's
eavesdropping will leave a trace in the results of the samples
$s_{e1}$, same as that in the high-dimensional BB84 QKD protocol
\cite{BP}. That is, the powerful eavesdropper, Alice will leave a
trace in the outcomes of the decoy photons as they are prepared with
some unbiased MBs randomly and Alice has no information about them,
including their positions and their states. Moreover, the
eavesdropping done by Alice in the quantum line between Charlie and
Bob can only obtain some information about the key $U_C$, not the
secret message $U_B$ at the risk of being detected. Charlie and Bob
can reduce the information leaked with quantum privacy
amplification.

In essence, the coding done by the users is performed only after
they confirm the security of the transmission in this QSDC network
scheme, same as the two-step QSDC protocol \cite{two-step} which is
secure as an eavesdropper cannot steal the information about the
secret message, similar to Bennett-Brassard-Mermin 1992 QKD protocol
\cite{BBM92,IRV,WZY}. The eavesdropping on the quantum line between
Alice to Charlie or Bob to Alice will be found out when Bob and
Charlie analyze the error rate of the outcomes of the Bell-basis
measurement on the EPR photon pairs on which Bob chooses randomly
one of the local unitary operations $U_{nm}$. And the eavesdropping
can only obtain the combination of the two operations $U_A= U_C U_B$
which will be announced in public.  In this way, the three
processes, Alice to Charlie, Charlie to Bob, and Bob to Alice, for
the transmission of the photons $S_B$ are similar to that between
the sender and the receiver in the two-step QSDC protocol
\cite{two-step}. If all the error rates are reasonably low, the
present QSDC network scheme can be made to be secure with some other
quantum techniques, such as quantum privacy amplification,
entanglement purification, quantum error correction, and so on.

As for the decoy photons, Charlie can also prepared them without
another single-photon source. He can produce them with measurements.
That is, Charlie requires Alice to choose $n_1+n_2$ photons in $S_A$
as the samples in the process for the first eavesdropping check.
Alice measures all these $n_1 + n_2$ photons by choosing randomly
the MB $Z_d$ or $X_d$, and publishes their states. But Charlie only
measures $n_1$ photons in his corresponding sampling photons in
$S_B$, and keeps the other $n_2$ photons as the decoy photons. As
Charlie completes the error rate analysis for the first
eavesdropping checking by himself, it is unnecessary for him to
publish the outcomes and  positions of the sampling photons measured
by Charlie. He need only publish the result that the error rate is
reasonable low or not. In this way, none knows which photons are the
decoy ones and their states. Also, Charlie can choose some unitary
operations to change the basis of a decoy photon. This good nature
will reduce the requirements on the users' devices as they need not
have an ideal single-photon source in their quantum communication.

As the quantum data in QSDC schemes
\cite{two-step,QOTP,Wangc,zhangzj,yan,gaot,zhangs} should be
transmitted with a block, the present QSDC network scheme with
quantum storage technique will have a high source capacity as each
photon can carry $2$log$_2d$ bits of information. Even though the
this technique is not fully developed, it is believed that this
technique will abe available in the future as it is a vital
ingredient for quantum computation and quantum communication and
there has been great interest \cite{storage} in developing it. As
only a sequence of photons $S_B$ is transmitted in the present QSDC
network scheme, it is more convenient than others \cite{lxhnetwork}.
The users need only exchange a little of classical information for
checking eavesdropping and the server publishes a bit of classical
information for each qubit, the present scheme is an optimal one
from the view of information capacity exchanged.

For implementing this QSDC network, high-dimensional quantum systems
are required. At present, people can choose polarized photons as the
two-dimensional quantum systems for this QSDC network without
difficulty. Another feasible candidate of quantum information
carriers may be the phases of photons. Recently, people employ
orbital angular momentum states of photons to carry more information
for communication \cite{OAM}. With the improvement of technology,
the generation and manipulation of high-dimensional quantum systems
become convenient and their application to optical communications is
feasible.

In summary, a scheme for quantum secure direct communication network
is proposed with superdense coding and decoy photons. After
confirming the security of the transmission, the user on the network
codes his secret message on the quantum states which has been
encrypted by the receiver with one of the local unitary operations
$\{U_{nm}\}$ chosen randomly. For preventing the dishonest server
from eavesdropping, the receiver picks up some samples from the
sequence received and replaces them with the decoy photons which can
be produced with measurements and unitary operations before he sends
the sequence to the sender. An authorized user on the network can
communicate another one securely. This scheme has the advantage of
high capacity as each photon can carry $2$log$_2d$ bits of
information and it is more convenient than others \cite{lxhnetwork}
as only a sequence of photons is transmitted in quantum line.

This work is supported by the National Natural Science Foundation of
China under Grant Nos. 10604008 and 10435020, and Beijing Education
Committee under Grant No. XK100270454.


\begin{thebibliography}{99}
\bibitem{note} If the users on a passive optical network do not
require that the severs provide the service for preparing the
quantum signal, all the point-to-point QSDC protocols existing can
be used to accomplish the secure communication on network. However,
the communication, in this time, is just a point-to-point one, and
each user should have the devices for preparing quantum signal and
even taking joint multipartite measurements.
\end{thebibliography}
\end{document}